\documentclass{article}
\usepackage{authblk}
\usepackage{hyperref}

\usepackage{graphicx}
\usepackage{bm}
\usepackage{float}
\usepackage{tabularx}
\usepackage{xfrac,cite}

\usepackage{verbatim}

\usepackage{xcolor}

\begin{document}                  
\setcounter{footnote}{1}
\title{Multi-Resolution Electron Spectrometer Array for Future Free-Electron Laser Experiments}


\author[1]{Walter, Peter\thanks{Electronic Mail: pwalter@slac.stanford.edu}\thanks{Authors contibuted equally.}}
\author[1]{Kamalov, Andrei \protect\setcounter{footnote}{2}\thanks{Authors contibuted equally.}}
\author[1]{Gatton, Averell}
\author[2]{Driver, Taran}
\author[1]{Bhogadi, Dileep}
\author[1]{Castagna, Jean-Charles}
\author[1]{Cheng, Xianchao}
\author[1]{Shi, Hongliang}
\author[1,b]{Cryan, James}
\author[5]{Helml, Wolfram}
\author[3,4]{Ilchen, Markus}
\author[1, 2]{Coffee, Ryan N.}

\affil[1]{SLAC National Accelerator Laboratory, 2575 Sand Hill Road, Menlo Park, Calfornia 94025, {USA} }
\affil[2]{The Stanford PULSE Institute, 2575 Sand Hill Road, Menlo Park, CA 94025, {USA}}
\affil[3]{European XFEL GmbH, Holzkoppel 4, 22869 Schenefeld, {Germany}}
\affil[4]{Institut f{\"u}r Physik und CINSaT, Universit\"at Kassel, Heinrich-Plett-Stra\ss e 40, D-34132 Kassel, {Germany}}
\affil[5]{Technische Universit\"{a}t Dortmund, Maria-Goeppert-Mayer-Str. 2, 44227 Dortmund, Germany}

\maketitle                        
\section{Abstract}
We report the design of an angular array of electron Time-of-Flight (eToF) spectrometers intended for non-invasive spectral, temporal, and polarization characterization of single shots of high-repetition rate, quasi-continuous, short-wavelength Free-Electron Lasers (FELs) such as the LCLS-II at SLAC. 
This array also enables angle-resolved, high-resolution eToF spectroscopy to address a variety of scientific questions of ultrafast and nonlinear light--matter interaction at FELs.
The presented device is specifically designed for the Time-resolved atomic, Molecular and Optical science end station (TMO) at LCLS-II.  
In its final version, it can comprise of up to 20 eToF spectrometers aligned to collect electrons from the interaction point defined by the intersection of the incoming FEL radiation and a gaseous target.
There are 16 such spectrometers forming a circular equiangular array in the plane normal to x-ray propagation and 4 spectrometers at 54.7$^\circ$ angle relative to the principle linear x-ray polarization axis. 
The spectrometers are capable of independent and minimally chromatic electrostatic lensing and retardation in order to enable simultaneous angle-resolved photo-electron and Auger electron spectroscopy with high energy resolution.
They are designed to ensure energy resolution of 0.25~eV across an energy window of up to 75 eV which can be individually centered via the adjustable retardation to cover ranges of electron kinetic energies relevant to soft x-ray methods, 0--2~keV.
The full spectrometer array will enable non-invasive and online spectral-polarimetry measurements, polarization-sensitive attoclock spectroscopy for characterizing the full time--energy structure of even SASE or seeded LCLS-II pulses, and also supports emerging trends in molecular frame spectroscopy measurements. 

\section{Introduction\label{sec:intro}}

Electron time-of-flight spectroscopy has substantially evolved over the past few decades toward providing high-resolution, high collection efficiency, and angle-resolved spectroscopic insights into a broad variety of scientific questions in different sample environments at both pulsed and continuous light sources. 
The instrument presented here is optimized for angle-resolving electron time-of-flight spectroscopy in the gas phase specifically focusing on the challenges associated with high repetition rate X-ray Free-Electron Lasers (XFELs).
Several instrumental advancements were needed to ensure high kinetic-energy resolution with moderate collection efficiency for the electrons under conditions where tens to hundreds of electrons are detected from a single x-ray pulse. 
From the first dedicated gas-phase eToF \cite{White79} to high-resolution and even angle-resolving eToF-spectrometers \cite{Becker89, Hemmers1998, Berrah1999}, advances over the past two decades have led to the first applications of this technique to exploring nonlinear \cite{Braune2016, Ilchen2018} and, recently, also time-resolving phenomena at FELs \cite{Hartmann2018}. 
A complete historic overview is not within the scope of this paper, but the mentioned milestones are intended to sketch with selected references the direct origins of the presented instrument.

The innovative concept of using an array of more than 10 independent time-of-flight spectrometers aligned in the dipole plane was first pursued by the group of U. Becker. 
It had the goal to capitalize on high energy resolution while substantially increasing the angular resolution \cite{Ilchen2014, Braune2018}. 
The subsequent development in the group of J. Viefhaus modified that concept, aiming for higher collection efficiency and the possibility for high retardation potentials that could cover the large photon-energy range from 250--3000 eV available at the P04 beam line at the PETRA III synchrotron, DESY, Germany \cite{Viefhaus2013a}. 
This instrument, colloquially named the ``Cookiebox,'' was successfully applied to the free-electron lasers FLASH in Germany, FERMI in Italy, and the LCLS in the USA, and inspired the development of a similar device for the European XFEL in Germany \cite{Laksman2019}. 
Braune \textit{et al.} furthermore presented a conceptually related instrument for spectral and spatial diagnostics at FLASH \cite{Braune2018}. 
The instrument presented here focuses on application to non-invasive reconstruction of the exact time, energy, and polarization structure of single x-ray shots for time-resolving investigations at XFELs. 
Additionally, it will allow for simultaneous acquisition of high-resolution spectra across different kinetic-energy regimes as required for correlation-based photo- and Auger electron spectroscopy.

The biggest challenges for this instrument are set by the pioneering advent of the LCLS-II XFEL. 
Specifically, LCLS-II will feature a 4~GeV continuous-wave superconducting LINear ACcelerator (LINAC) that will produce equidistantly and potentially variably spaced ultrashort x-ray laser pulses with repetitions rates up to 1 MHz.
In the soft x-ray regime, it will span a photon energy range from 0.25 keV to 5 keV. 
The superconducting LINAC can deliver x-rays with sub-femtosecond duration via XLEAP \cite{Duris2019}, multi-color pulses \cite{Lutman13_twocolor,Durr2016}, and after installation of the afterburner quadrupole DELTA undulators will allow full polarization control\cite{Lutman2016} including time-dependent polarization \cite{Sudar2020}. 
Seeded operation and a variety of pump--probe schemes will further enrich the versatility of this new milestone of XFEL-developments.
Control over and complete characterization of these shot properties are critical for the bulk of the visionary science cases of LCLS-II \cite{Schoenlein2015} that target pressing contemporary challenges in ultrafast and nonlinear spectroscopy.

The array of time-of-flight spectrometers presented here, the so-called Multi-Resolution ``Cookiebox'' Optimized for the Future of Free-Electron (laser) Experiments - ``MRCOFFEE,'' aims to provide an XFEL-optimized single-shot instrument that targets the full time--energy reconstruction with attosecond temporal resolution as per Ref.~\cite{Hartmann2018} with (time-dependent) polarization reconstruction as per Ref.~\cite{Lutman2016}.
Targeting also attoclock spectroscopy, we have additionally designed for appropriate collection efficiency across multiple degenerate-positioned spectrometers which can be operated with independently controlled retardation fields in order to simultaneously measure the time--energy emission pattern at multiple photoionization and Auger-electron emission energies.
The system is designed specifically for compatibility at the ultimate 1~MHz repetition rate of LCLS-II such that it enables both attosecond scale time-dependent experiments \cite{Taran2020} and few-femtosecond scale nonlinear chemical experiments \cite{RostPRL2020}, as expected to rapidly evolve in the coming years of FEL science.

A principle strength of the spectrometer array lies in the very high time resolution of the Micro-Channel Plate (MCP) based detectors.
The MCP detectors record electron counts from single pulses at high repetition rates with excellent time resolution and narrow signal pulse envelopes, here 450 ps (FWHM).
This narrow pulse width allows for the time resolution necessary to preserve the stringent energy resolution for the very short flight times associated with the upper end of the pass energy window.
Alternatively, the MCP detectors can also be run in analogue mode whereby the current through the MCPs is measured.
This analog mode, as opposed to counting mode, can be used to accommodate conditions where single-electron counting is either prohibitively difficult or not desirable, both of which are frequently the case at XFELs.

\section{Hardware\label{sec:hardware}}

As introduced above, this diagnostic and experimental instrument, MRCOFFEE, will record electron spectra over independent energy ranges with high resolution for ultra-intense XFEL pulses.
This section describes the engineering challenges required for such a tailored instrument and is organized by component: the main chamber, the spectrometers, and the target delivery and x-ray alignment tool.
\subsection{The main chamber}
The main chamber is engineered to house 16 eToF spectrometers orientated azimuthally around the x-ray trajectory equally spaced by 22.5\textdegree.
There are an additional 4 spectrometers oriented at the so called ``magic'' angle of 54.7$^\circ$ relative to the horizontal x-ray polarization.
This particular angle records a signal that is independent of angular emission pattern and therefore proportional to total yield \cite{Kivimaki1998} and aids the retrieval of laboratory frame angular asymmetries in electron emission patterns \cite{Catalin2013}. 
The chamber has additional ports for four non-magnetic bearing turbomolecular vacuum pumps, sample delivery, x-ray beam diagnostics, vacuum feedthroughs, and viewports. 

A common spectrometer centering ring mounted at the chamber center allows for precise mechanical alignment of all spectrometers to a common interaction region with better than 100~$\mu$m diameter intersection volume.
This precise mechanical alignment, as with previous designs, relieves the need for independent manipulators for each spectrometer.

\begin{figure}
    \centering
    \includegraphics[clip,width=.8\linewidth]{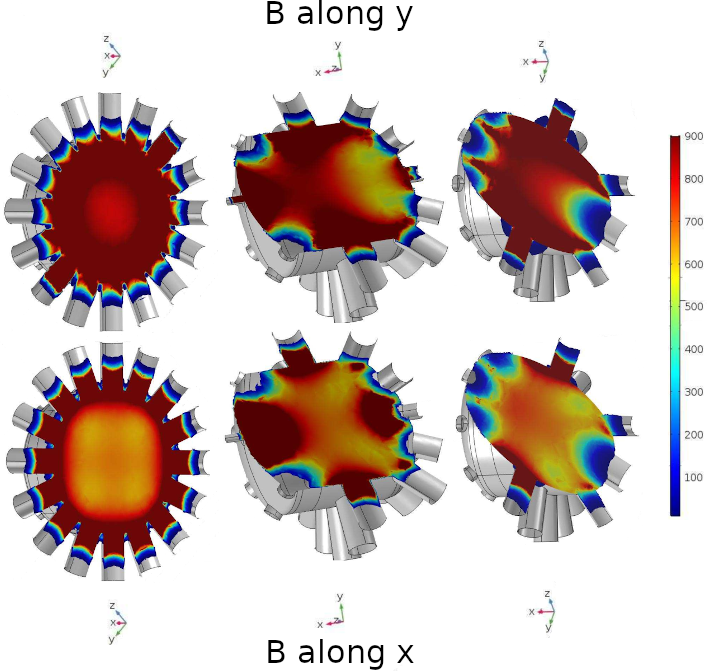}
    \caption{
    Magnetic shielding factor (attenuation factor) calculations for the final experiment chamber design.
    Magnetic attenuation calculation performed by Scientific Magnetics \cite{scimagnetics}.
    }
    \label{fig:MagShielding}
\end{figure}

Electrons that encounter non-homogeneous magnetic fields in flight to the MCP detectors would distort the yield and resolution of low-kinetic-energy electrons. 
To avoid the effects of such pathological steering, the main chamber is manufatured of high magnetic permeability mu-metal, as are the spectrometer attachment spool pieces (see Sec.~\ref{TOF}).
The mu-metal construction of the chamber and spools, along with mu-metal bridging rings that facilitate the magnetic transition across the stainless flanges, significantly reduce the latent magnetic field in the flight path of the spectrometers (see figure \ref{fig:MagShielding}). 

\begin{figure}
\centerline{a\hspace{.5\linewidth}b}
\centerline{
\includegraphics[trim=50 0 50 0,clip,width=.5\linewidth]{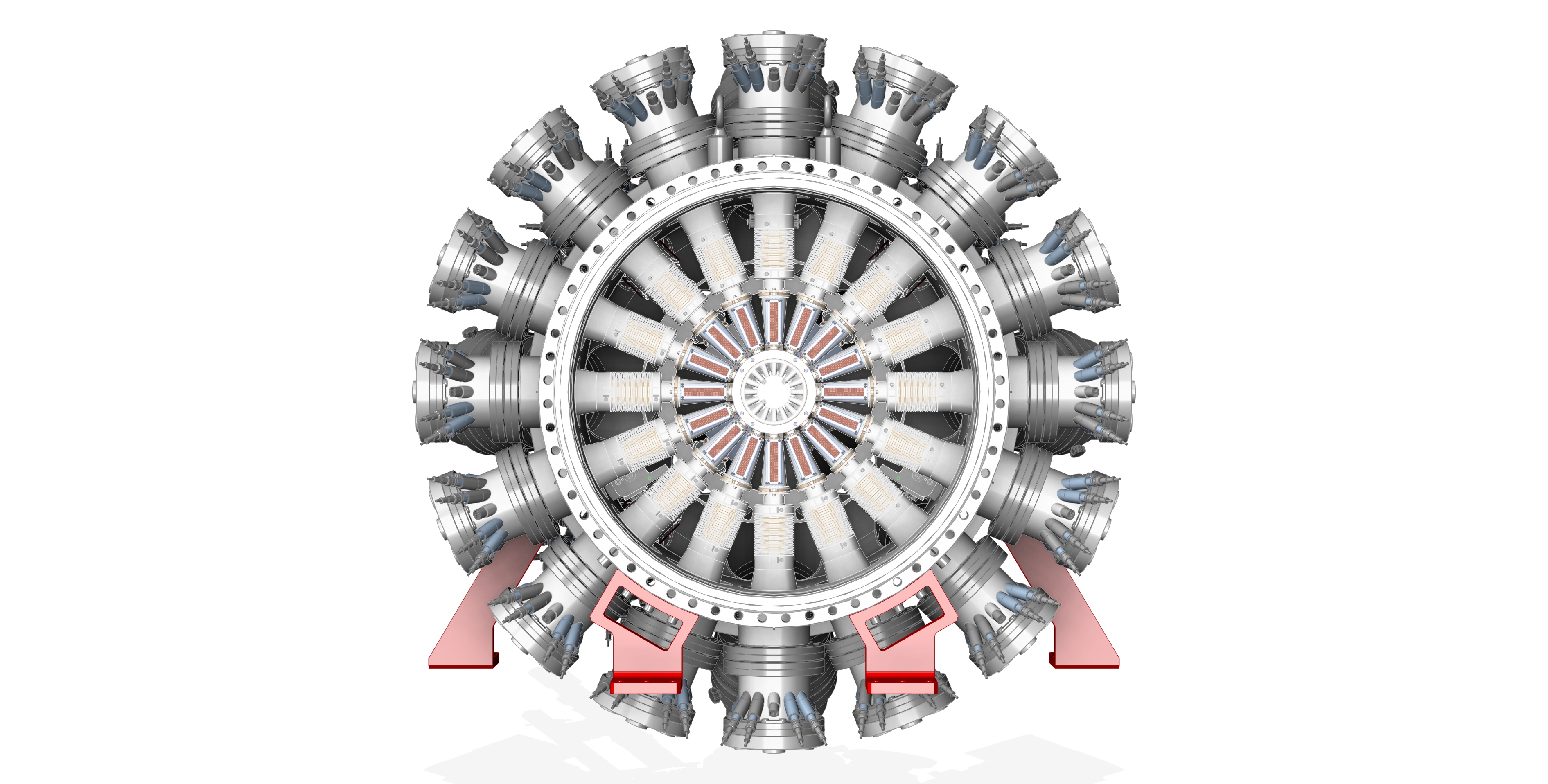}
\includegraphics[trim=50 5 70 5,clip,width=.5\linewidth]{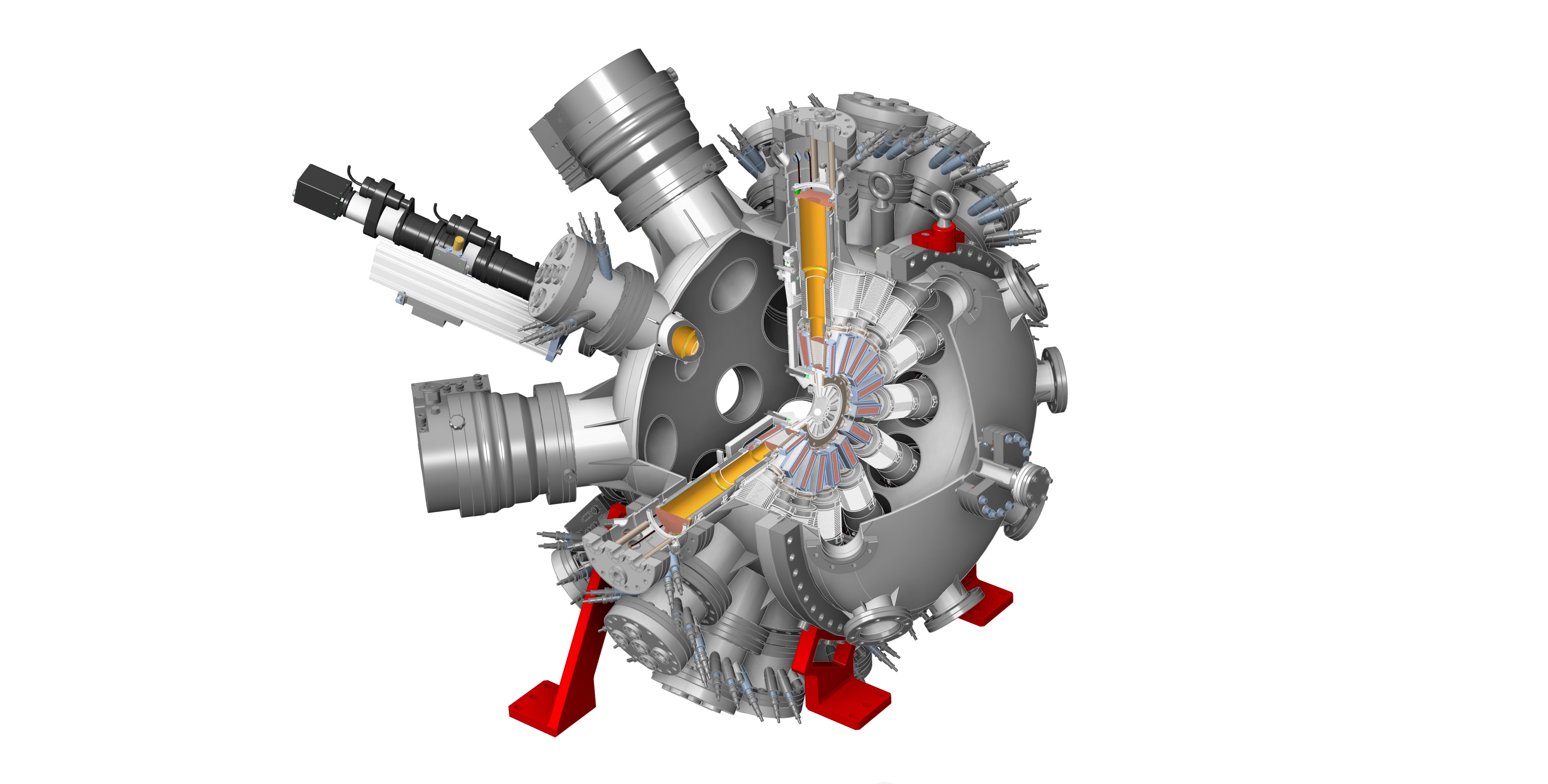}
}
\caption{
a) Axial view of the detector array along the direction of x-ray propagation, showing the 16 azimuthal spectrometers. 
b) Section view partly showing ``magic'' angle spectrometers, turbomolecular pumps, and an interaction viewing camera.
}
\label{fig:CH1}
\end{figure}

We avoided magnetic materials such as nickel and stainless steel for internal components whenever possible to ensure high collection efficiency across a 75~eV resolvable energy window.
The main chamber schematic is shown in figure \ref{fig:CH1} and the expected magnetic shielding is shown in figure \ref{fig:MagShielding} for magnetic fields ($\vec{\bm{B}}$) vertically perpendicular to the x-ray trajectory (top row) and for magnetic fields horizontally perpendicular to the x-ray trajectory (bottom row). 

\subsection{The spectrometers}\label{TOF}

Given the number of spectrometers in the array, we require a reliable eToF spectrometer design that can provide high resolution and collection efficiency at any user specified energy region.
Using a uniform modular design ensures serviceability and allows reconfiguration or replacement for selectiv angular configurations. 
All spectrometers operate simultaneously to measure angle-resolved electron spectra as per the target applications of attosecond angular streaking measurements, LCLS-II diagnostics, and a broad range of scientific measurements.

\begin{figure}
\centerline{
\includegraphics[width=0.95\textwidth]{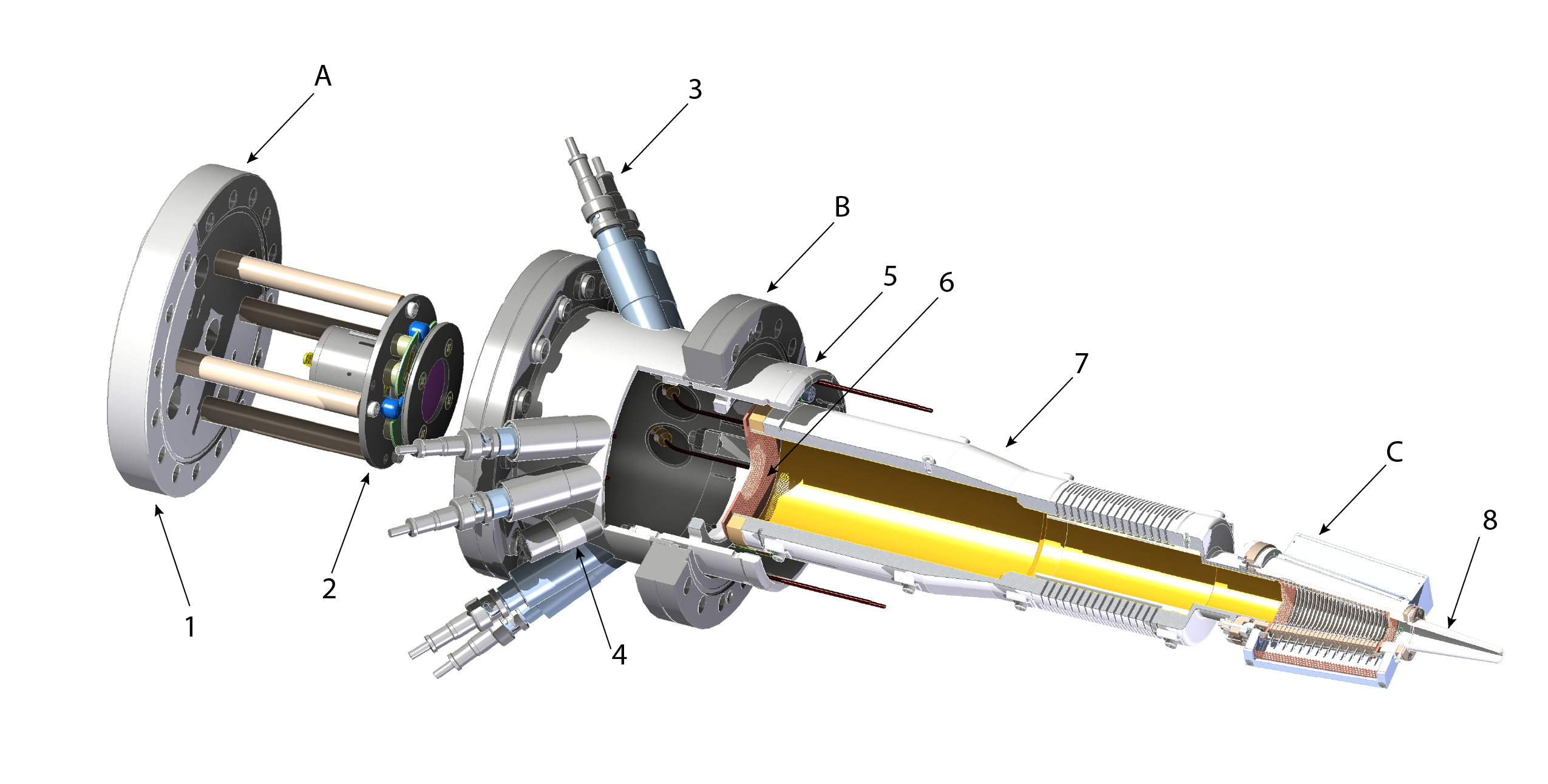}
}
\label{fig:TOFa}
\caption{The spectrometer. 
Shown are (A) the MCP assembly, (B) the flight-tube assembly, and (C) the lens-stack assembly (see figure \ref{fig:lensStack} for detailed view). 
The MCP assembly consists of a (1) feedthrough flange with 5 mini-flange openings for individual feedthroughs and (2) the fast response MCP-detector. 
The drift-tube assembly consists of (3) 8 high-voltage feedthroughs, (4) two 9-pin high-voltage multi-pin feedthroughs, (5) a mu-metal bridge, (6) the end-of-drift-tube mesh, and (7) the drift tube. 
(8) shows the flight-tube nose cone. }
\end{figure}

Our spectrometer design is presented in detail in figure \ref{fig:TOFa} with distances and acceptance diameter summarized in Table~\ref{tab:geometry}.
Entering electrons first encounter a series of electrostatic lenses which retard and partially collimate the electrons to improve the energy resolution.
Electrons then traverse a gradient free drift tube and progress through a meshed aperture at the exit of the drift region.
Immediately upon exiting the flight tube, electrons are pre-accelerated into a commercially available MCP detector (Hamamatsu model F9890-31(-32)).
The spectrometer assembly is mounted in the mu-metal spool piece that supports the lensing section, and the drift tube assembly.
The MCP detector is separately attached to be re-entrant into the mu-metal spool such that it benefits the magnetic shielding but remains separately mounted for ease of replacement, reconfiguration and service. 
The mu-metal spool piece is equipped with ten High-Voltage (HV) feedthroughs, eight that are compatible with 6~kV low-current supply and two nine-pin feedthroughs that are compatible up to 1~kV. 
The total of 26 HV feedthroughs allows for future development of more finely controllable electrostatic potentials.
The drift tubes incorporate a flexxible section in order to accommodate thermal expansion during vacuum bake and also to ease the mechanical register of the spectrometer nose section into the socket of the centering ring. 
\begin{table}

\caption{
\label{tab:geometry}
The four principal regions of the spectrometer, defined by the distance from the x-ray/target interaction point.
Relative solid angle acceptance is noted in \% of a unit sphere.
\protect\footnotemark Slightly larger opening of 0.220 \% than upstream acceptance limitation.
\protect\footnotemark Current MCP diameter is 27 mm giving 0.029\% angular acceptance.  
}
\addtocounter{footnote}{-2}
 
\begin{tabular}{l|cccc}
Region & Start [mm]& Stop [mm] & Stop Aperture & rel. solid angle\\
        &           &       & Diam. [mm]  & [\%]\\
    \hline
Interaction & 0.0 & 25.0 & 4.36 & 0.189\\
Nose & 25.0 & 80.0 & 15 & 0.180\\
Lens/retardation & 80 & 132.8 & 25 & 0.220\footnotemark\\
Drift & 125 & 396.3 & up to 57\footnotemark & up to 0.129 
\end{tabular}
\end{table}

\subsection{The lens stack}\label{LENS}   
\begin{figure}
\centering
\includegraphics[width=0.9\textwidth]{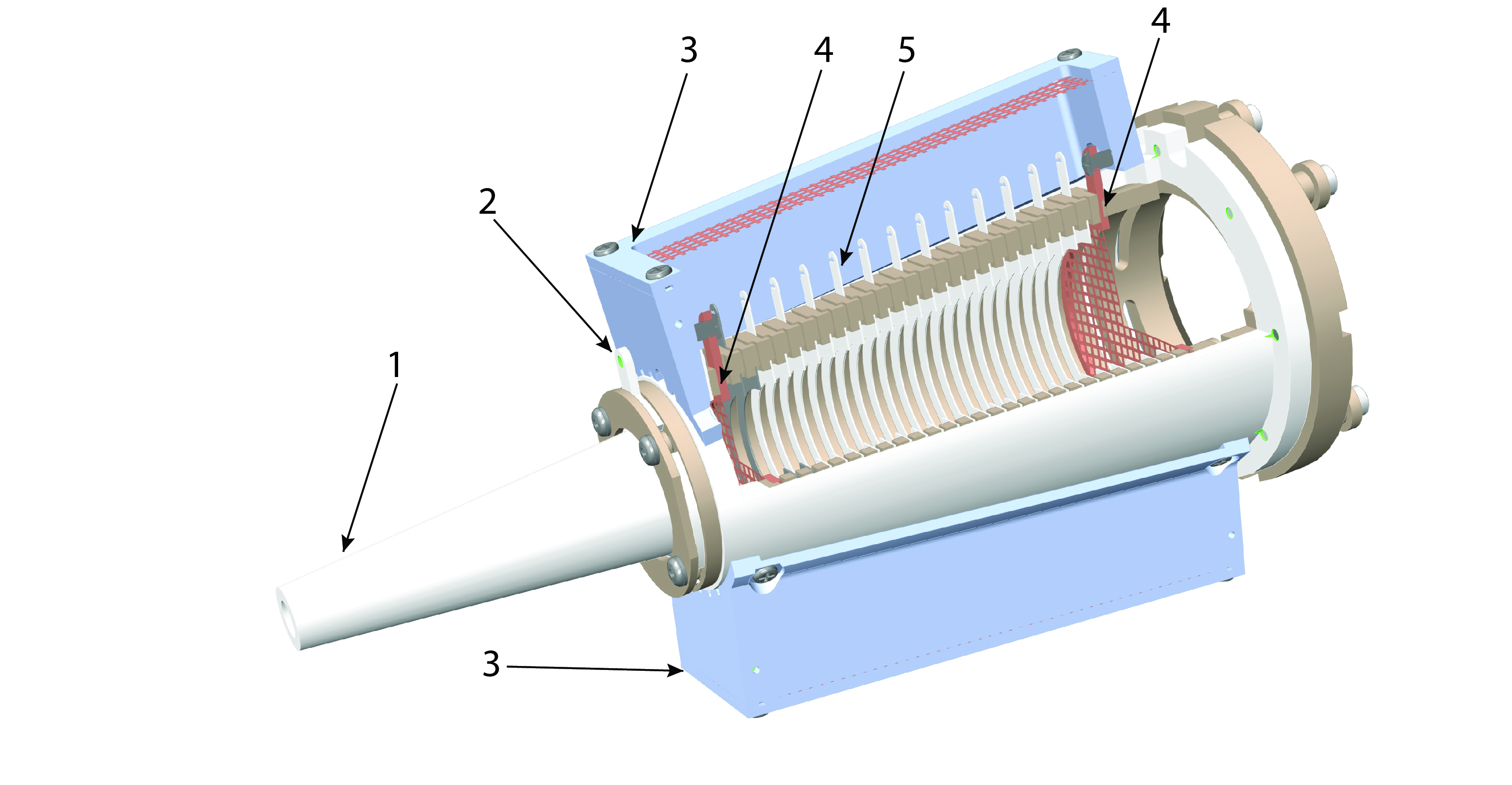}
\caption{
Cross-section of the electrostatically retarding compound lens. 
Shown are (1) the nose cone and respective (2) HV connection, (3) RF shielding cages for the resister bridge, (4) copper meshes at the beginning and end of the retardation lenses and (5) the stack of 25 lensing blades.
}
\label{fig:lensStack}
\end{figure}

\begin{figure}
\centering
a.\hspace{.5\linewidth}b.\\
\includegraphics[clip,height = .3\linewidth]{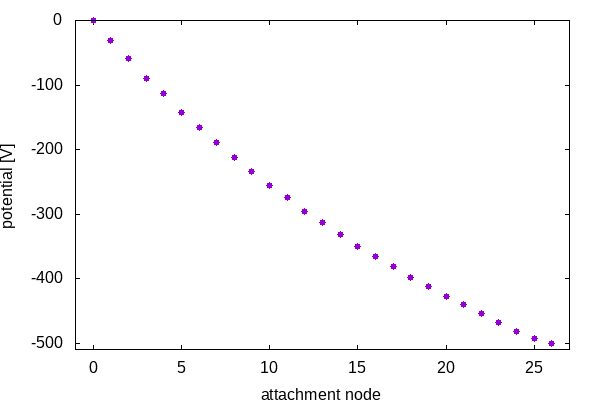}\hfill
\includegraphics[clip,height = .3\linewidth]{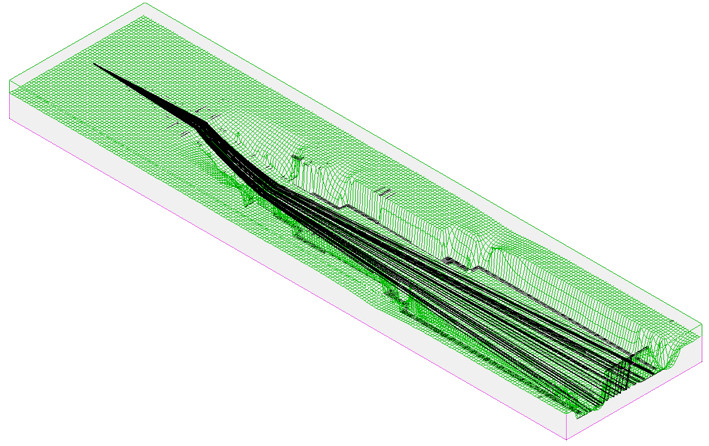}\\
c.\hspace{.5\linewidth}d.\\
\includegraphics[clip,height = .3\linewidth]{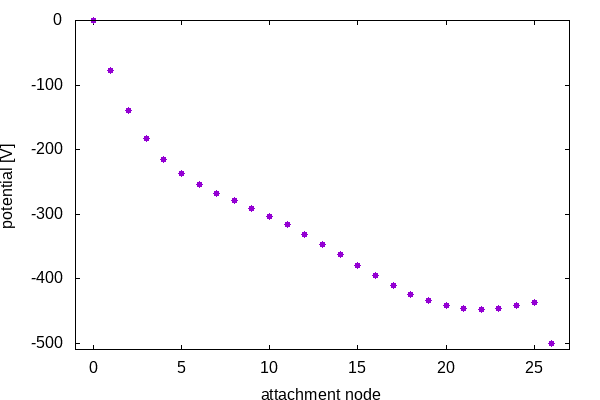}\hfill
\includegraphics[clip,height = .3\linewidth]{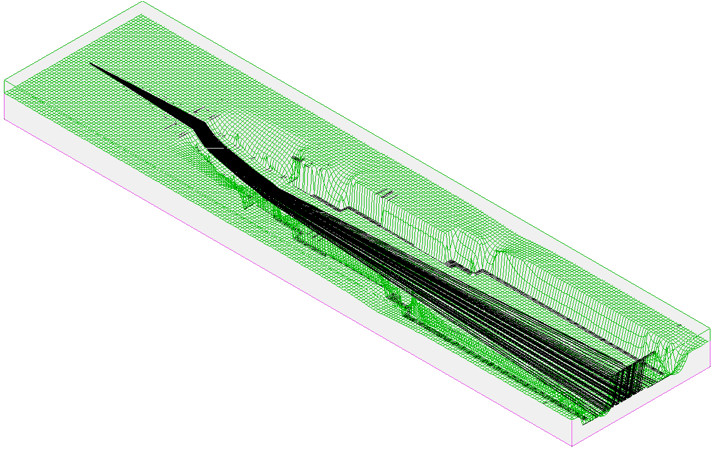}
\caption{
(a) $\sfrac{1}{r}$ potential field that retards electrons without altering the angle of (b) the corresponding trajectories.
(c) Optimized potential profile and (d) the corresponding trajectories.
}
\label{fig:retardation}
\end{figure}

The electron collection of the individual spectrometers is controlled by a series of 25 copper blades positioned at the front of the spectrometer modules as shown in figure~\ref{fig:lensStack} that form a compound electrostatic lens with energy retardation.
The blades are inter-connected via a resistor bridge which, combined with the supplied voltages, determines the field gradient via the voltage set points of the individual blades, denoted as ``nodes'' in figure~\ref{fig:retardation}a and c.
These nodes correspond to the entrance mesh at node 0, blades from 1--25, and exit mesh at 26.
The resistor-bridge values were selected to maximize collection efficiency while preserving energy resolution.
The electric potential steadily drops (increasingly negative) throughout the blade stack, lowering the kinetic energy of electrons to increase the respective energy resolution.
The final part of this compound lens uses a sudden positive bump intended to better collimate the electron trajectories onto the MCP detector.
Fine copper meshes with a transmission greater than 90\% are positioned at the front and back of the compound lens.
This ensures that the electrostatic potential is flat outside of the lens stack for free propagation to the MCP detector.
The performance of the lens design was simulated using SIMION and is shown in figures~\ref{fig:resolutionComparison} and \ref{fig:collectionEfficiency}.

The design of the spectrometer allows for different modes of operation. 
One principle and high profile use for the MRCOFFEE array is attoclock spectroscopy which requires a wide energy acceptance range of $10-75$~eV above the nominal retardation while maintaining $0.25$~eV resolution.
Given the number of HV feedthroughs, nearly every blade can be supplied by its own HV source, supplied at the nodes in figure~\ref{fig:retardation}a and c.

To achieve optimal resolution, at the expense of collection efficiency, we first investigated a $\sfrac{1}{r}$ potential profile.
Under this condition, as can be seen in figure \ref{fig:retardation}b, particle trajectories that emanate radially from the interaction volume largely continue through the retardation section with minimal transverse deflection.
This configuration preserves high energy resolution but begins to sacrifice collection efficiency, especially at larger energy retardation settings.
In this case, we define the total retardation potential by setting the mesh and drift region potential both to $V_{\mbox{ret}}$, while holding the front mesh and nose to ground potential.

A preferred alternative to the $1/r$ profile improves collection efficiency while minimally impacting energy resolution by using the compound transverse lensing action to direct a greater number of trajectories onto the MCP.
We used the Nelder-Mead optimization algorithm with an objective function that optimizes toward a monotonic mapping of the acceptance solid angle to the 27~mm diameter MCP detector area.
This optimization produced a compound lens with a final collimating sub-lens that collimates the trajectories at the end of the blade stack to traverse onto the MCP, as shown in figure~\ref{fig:retardation}d.

\begin{figure}
\centering
a.\hspace{.5\linewidth}b.\\
\includegraphics[clip,width=.9\linewidth]{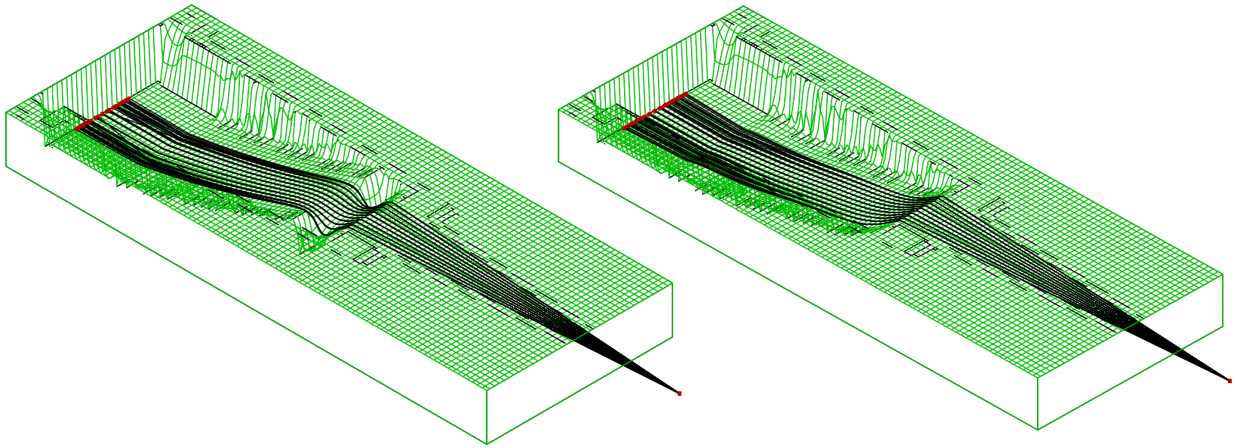}
\caption{
(a) Trajectories in the electric field for a conventional electrostatic lens at the entrance and blade stack regions.
(b) Trajectories for an optimized compound electrostatic lens.
}
\label{fig:focus}
\end{figure}

We compare a conventional electrostatic lens design with that presented here as a compound retarding lens in figure.~\ref{fig:focus}.
As can be seen in panel a, the outermost trajectories are over-bent back inward and will cross over the center line in the drift region (not shown).
This effect will lead to a region of high-density hits which may saturate the MCP detector and decrease the collection efficiency of the spectrometer for particular pass energies. 
This is the well known chromatic ``sausage'' effect whereby some energies are over focused, others are focused into the center of the detector, and yet higher energies are under-focused.
This is of particular concern for the high-repetition-rate LCLS-II since the amount of electrons delivered in a small focused spot over time could be potentially damaging for a MCP.
We therefore prefer to ensure that no energies will be tightly focused into the center of the MCP.

This motivates our search for a strictly monotonic mapping from electron emission angle to radial detection location at the MCP.
The Nelder-Mead optimization algorithm was constrained with this goal in mind.
Figure~\ref{fig:focus}b shows that the optimized compound lens collimates an appropriately monotonic series of trajectories similarly to the conventional case yet none of the trajectories explicitly cross.

\subsection{Gas delivery \& interaction alignment}
\begin{figure}
a\\
\centerline{
\includegraphics[trim={0 75 0 25},clip,width=.9\linewidth]{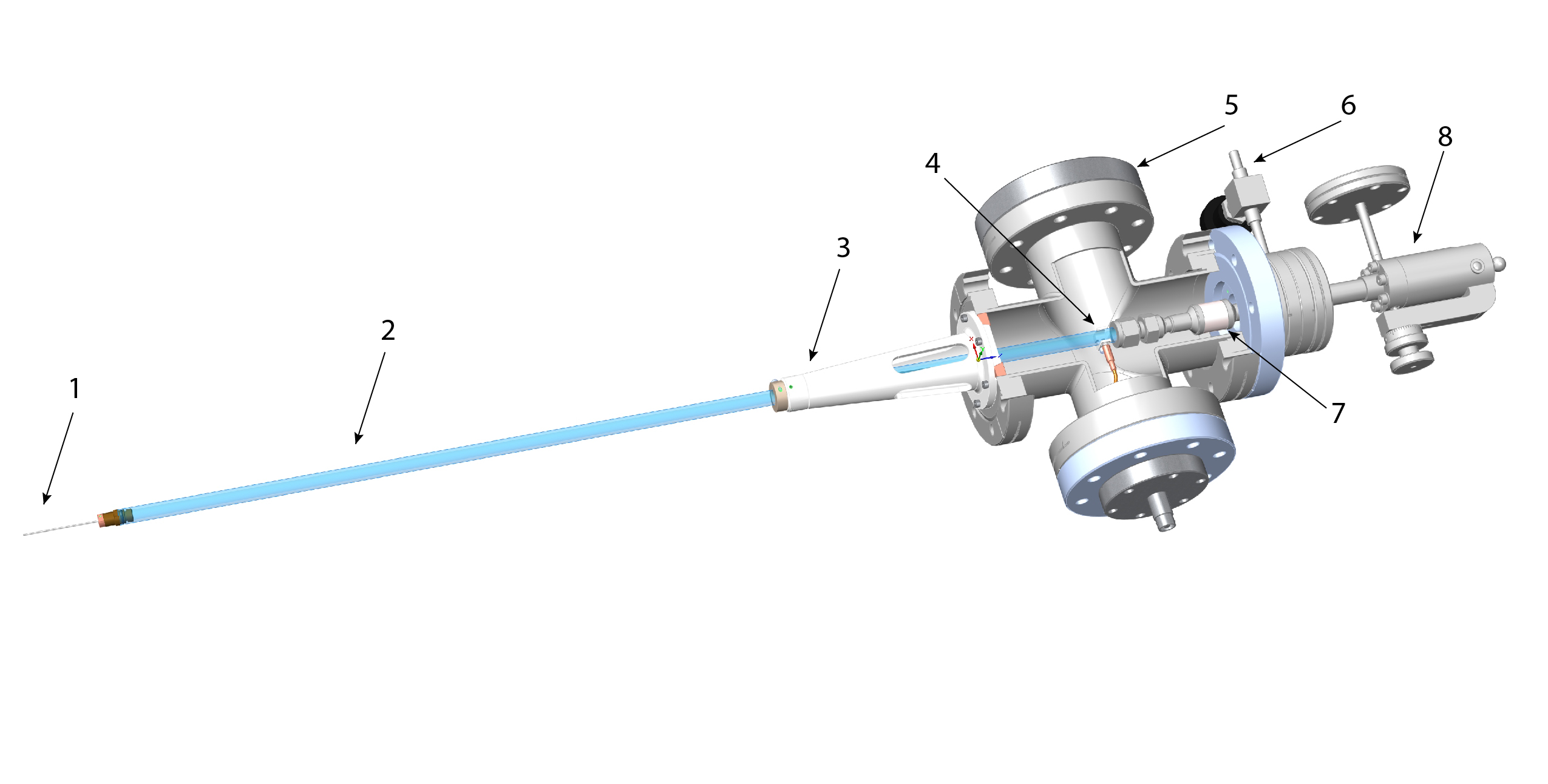}
}
b\\
\centerline{
\includegraphics[trim={0 0 0 25},clip,width=0.9\linewidth]{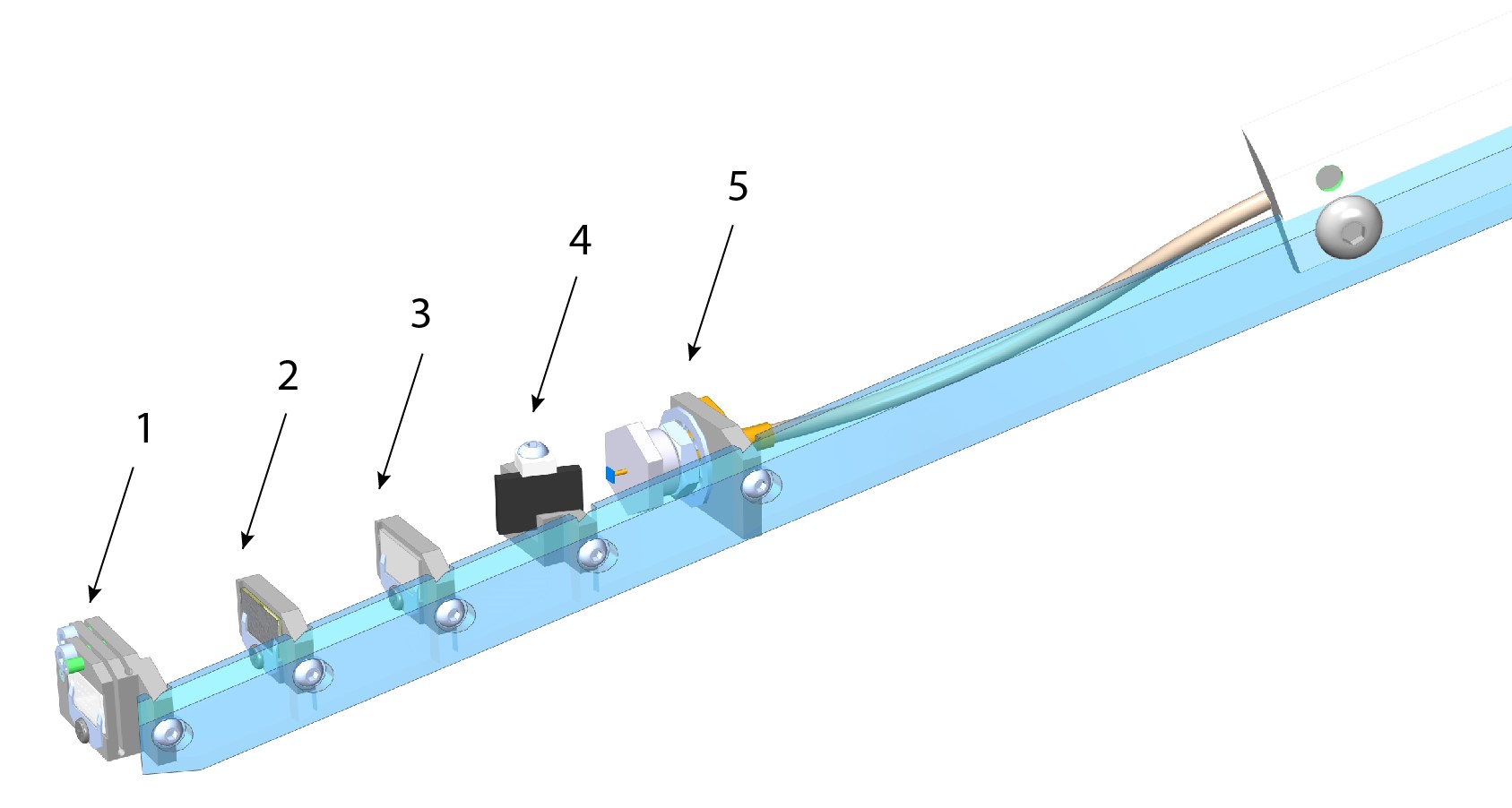}
}
\label{fig:DIA}
\caption{
a. Gas-sample injection assembly. 
Shown are the (1) sample gas nozzle, (2) the gas support tube, (3) the alignment stabilizer, (4) HV connection, (5) additional feedthrough flange, (6) pump and purge port, (7) electric isolating inlet coupling, and (8) the leak valve.
b. Interaction alignment and x-ray focus diagnostic paddle. 
Depicted are the (1) flexture holder for the pinhole array, (2) $\&$ (3) the YAG and filter holders, (4) the optical laser scatter target and (5) the timing diode.
}
\end{figure}

The main sample delivery apparatus is an effusive gas nozzle pressurized via a flow controller. 
The sample delivery rod can be set to high voltage in the event of charging and also to accomodate a deliberate need to apply an electrostatic field. 
The in-vacuum interaction end of this rod is insulated from the gas inlet source end with fields applied via a HV feedthrough at the source cross. 
Two sub-D feedthroughs can be used to install a sample heating line up to the tip (see figure \ref{fig:DIA}). 
For precise alignment, the sample delivery apparatus is mounted on a motorized three-way manipulator.  \\

The instrument is equipped with a multi-purpose interaction diagnostic paddle, mounted on a three-way manipulator, to accommodate x-ray focus diagnostics and beam positioning as well as coarse x-ray arrival time determination.
For commissioning, calibration, and alignment of the TMO wavefront sensor, the first position of the diagnostic paddle (1) is a three-axis flexture that allows alignment of a pinhole array wafer perpendicular to the x-ray propagation vector. 
Positions (2) and (3) host $10\times5$~mm wafers such as YAG, imprint samples, and/or filters that can be individually selected depending on the experiment and user needs. 
Position (4) is a fluorescence target used with a high-speed photodiode (5) for coarse x-ray timing; paddle motion allows for both options, fluorescence or hitting the diode directly. 

\section{Preliminary results\label{sec:prelimresults}}

A principle challenge for MCP-based instruments such as presented here lies in the analogue mode of the MCP detectors as discussed above.
The tolerable number of electrons per shot is one limit, but for high repetition rate XFELs the longevity of the MCP over the longer time span of weeks and months becomes also a concern. 
The latter is principally set by the charge replenishment rate of the MCPs while the former is a function of the MCP aging.
in order to provide 
One way to mitigate these effects, and to properly ensure sufficient re-charge of the micro channels for constant intra- and inter-shot MCP gain is to run at the minimum potential across the MCP stack that preserves acceptable, but potetntially not optimal, signal-to-noise and then use matched bandwidth amplification ideally as close to the anode as possible.
For this reason, we plan for broad band 0.1--3 GHz signal amplification immediately at the signal output from the MCP detector flange.

\begin{figure}
    \centering
    \includegraphics[width=0.6\textwidth]{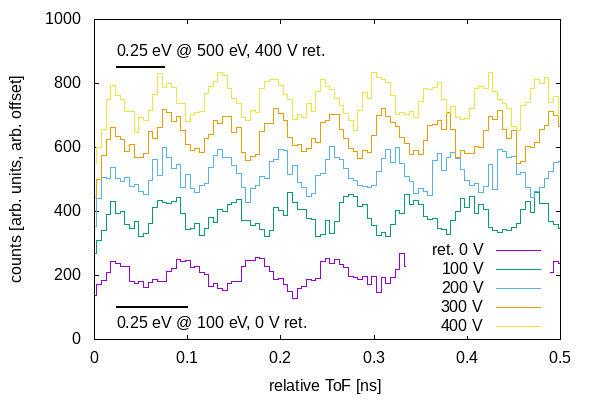}
    \caption{
    Simulated comparison of the time-of-flight resolution versus retardation.
    The flight time histograms are plotted relative to the first arrival of 100 eV pass energy electrons.
    These are followed by a series of subsequently lower energy peaks (later arrival) centered at 0.25~eV decremented energies with Gaussian distributions of $\sigma=0.0625$~eV.  
    }
    \label{fig:resolutionComparison}
\end{figure}

\begin{figure}
    \centering
    \includegraphics[width=0.6\textwidth]{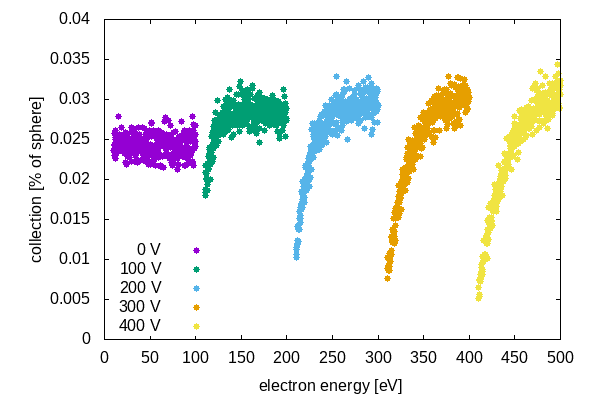}
    \caption{Simulated collection efficiency comparison for a single spectrometer at various electron energy ranges with corresponding retardation potentials.  
    Retardation is noted in the key and simulation results are shown for 10--100eV above each retardation potential.
    }
    \label{fig:collectionEfficiency}
\end{figure}

Another limitation concerns the tolerable electrons per shot that will avoid space charge induced degradation of the energy resolution that happens directly in the interaction volume.
We base our estimates for this limit on previous experience with thei original Cookiebox of Jens Viefhaus and its replica unit at the European XFEL \cite{Nick2018,Lutman2016,GregorRSI2016}.
In those experiments, we found that energy distortion occurred due to space charge at the level of 10$^4$ electrons produced in an interaction volume of 5 $\mu$m diameter \sfrac{1}{e}-spot by 100~$\mu$m Rayleigh range; this equates to about $4 e^- \mu\mbox{m}^{-3}$.
Depending on the particular use-case, we generally estimate a limit of about 10$^5$ $e^-$ generated in a 15 $\mu$m diameter \sfrac{1}{e}-spot of 100~$\mu$m, e.g. $\sim17\times 10^3~\mu\mbox{m}^3$ interaction volume. 
With an expected collection of near $3\times10^{-4}$ (see figure.~\ref{fig:collectionEfficiency}) we would expect this limit would provide about 30 collected electrons for each MCP detector.
We consider this an appropriate transition condition between the counting regime and the analog regime, of course dependent on the spread of the electron energies across the resolution window.
Figure~\ref{fig:resolutionComparison} shows, in simulation, that individual peaks split by 0.25~eV in the vicinity of 100 eV above the retardation potential require $e^-$ arrival time resolution that is well below the available digitizer sampling period of 0.166 ns.
For this reason, we rely on digitital signal processing as summarized in Sec.~\ref{sec:sigproc} to resolve hits with sub-sampling resolution.

Figure \ref{fig:collectionEfficiency} shows the fractional angular collection as a percentage of unit spherical emission.
We restrict the pass energy window to above 10~eV due to a preponderance of secondary electron emission below 10~eV that would otherwise obscure the desired target molecule spectra.
This simulation shows that the optimized field gradient retards high-energy electrons without significantly sacrificing collection.  
The absolute efficiency is set by the solid-angle size of an individual spectrometer's entry aperture and corresponds to a maximum possible collection percentage of 0.189\% as noted in Table~\ref{tab:geometry}.



\section{Signal Processing\label{sec:sigproc}}

The full full instrument allows for up to 20-fold parallel signal readouts that can in turn be used for real-time data sorting, veto, and statistical analysis of downstream detectors.
At its most simple form of veto, low latency results will allow users to choose on the fly only those x-ray shots with desirable pulse parameters; at its most advanced, semi-autonomous control could automatically drive the XFEL toward conditions that are likely to expose more information-rich regions of parameter space.
For example, in attosecond physics experiments \cite{Nick2018,Taran2020}, one could ultimately wrap a control loop around pump-probe delay settings in XFEL such that the experiment would self tune to information rich delay and multi-pulse energy settings. 
In polarization-based studies, one could use the time- and spectrum-resolved polarization results to aggregate like shots or even control the non-trivial shaped polarization states \cite{Sudar2020} to drive toward conditions that expose novel chiral dynamics.
At the very least, the instrument would serve as the first data reduction step for LCLS II and LCLS-HE experiments, providing a critical function for experiments that will soon incorporate 2D area detectors capable of serving the 1 MFps imaging rate.
We have co-designed the detector array along with analysis algorithms that will accommodate low-latency decisions based on x-ray pulse characteristics.
We will implement these algorithms in a data-flow paradigm using FPGA-based streaming analysis that divides x-ray pulse reconstruction into stages of increasing complexity.
Although the deeper layers of analysis, performed later in the data transformation chain, are beyond the scope of this instrumentation manuscript, we note that they presuppose a so-called ``featurized'' representation of the sigle shot electron energy spectrum.
It has been recently shown that an FPGA-implemented inference engine for course grained x-ray properties required the energy domain histogram rather than the direct time-of-flight spectrum \cite{Audrey2019}.
We plan to produce the energy domain histogram directly in the FPGA on the waveform digitizers themselves and so have chosen digitizers that allow for custom firmware [Abaco Systems Inc.]. 

Given the hit rate and space charge limits discussed in Sec.~\ref{sec:prelimresults}, we will adjust the target gas density to limit the incidence hit rate to below 100 counts per shot per MCP detector, ideally in the range 10--30.
We therefore focus on the counting regime for the MCP detector signals, leveraging the fast rising edge of electron cascade signal for attaining sub-sampling flight time resoluion.
We also reduce the total charge needed to register a hit and favor fast fall time as well since the individual spectra will likely be quite congested in this rather high counts rate regime.
For initial demonstrations, we have chosen the Hamamatsu F9890-31/32 detector assembly which supports a FWHM rise-time resolution below 0.5~ns.
We rely on external high-bandwidth signal amplification to accommodate reduced MCP gain in order to mitigate charge depletion while preserving sensitivity and rising edge bandwidth.

\begin{figure}
    a.\\
    \centerline{
    \includegraphics[clip,width=.6\linewidth]{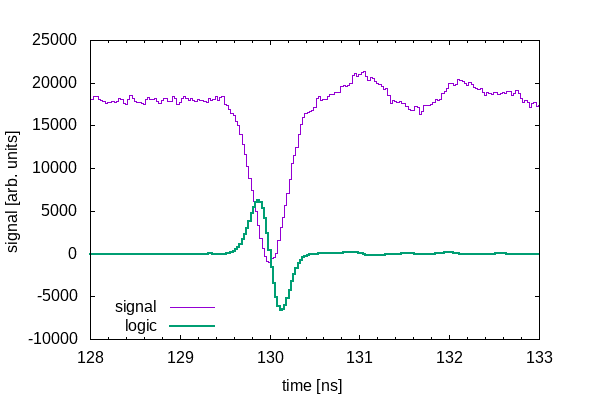}
    }
    b.\\
    \centerline{
    \includegraphics[clip,width=.6\linewidth]{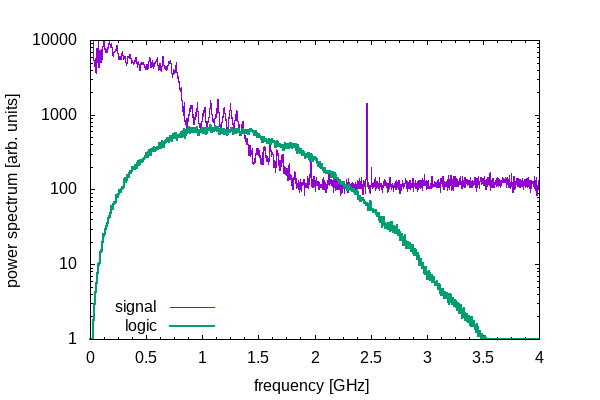}
    }
    c.\\
    \centerline{
    \includegraphics[clip,width=.6\linewidth]{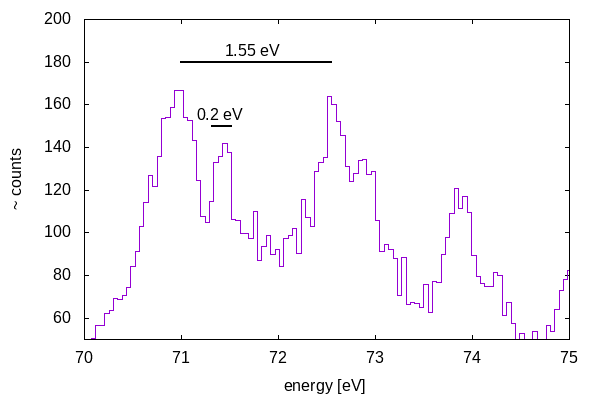}
    }
    \caption{
    a.~Waveform with an example signal amplifier minicircuits model CMA-83LN+ and the ``logic'' waveform as described in the text. 
    b.~The average power spectrum of the signal and logic waveforms.
    The logic waveform preserves the high frequency region of the signal but suppresses the low frequency and as well mitigates the residual post-pulse over shoot.
    c.~Measured Argon ATI spectrum with 70~V of accelerating potential applied.
    }
    \label{fig:waveformspect}
\end{figure}

In order to achieve the best resolution in the energy representation, we recognize that the precision in timing information is contained in the high frequency region of the signal power spectrum.
Figure \ref{fig:waveformspect} follows our signal processing algorithm whereby we read the amplified signal as discussed in Sec.~\ref{sec:prelimresults}.
The minor ticks in figure~\ref{fig:waveformspect} represent the expected digitization step size for the TMO instrument.
For the sake of testing, however, waveforms were measured with a 40 GS/s (20 GHz bandwidth) LeCroy digital oscilloscope.
We optimized the bias voltage across the MCP in order to optimize the signal to noise ratio for frequencies above 1~GHz.



accept only shots within a user-specified pulse duration.

\section{Diagnostic and Scientific application\label{sec:diagsciapps}}



The experimental paradigm of ``measure-and-sort'' has grown ubiquitous to the XFEL community.
Be it related to the x-ray arrival time diagnostics \cite{MinaRSI}, the numerous electron bunch diagnostics, or the spectral properties of the x-ray pulses themselves \cite{Taran2020}, users have become expectant of and profited significantly from single-shot analysis for sorting and other algorithmic treatment of the science data accordingly for each unique XFEL shot.

The instrument presented here is expected to serve this paradigm by providing single-shot reconstruction of the full multi-color pulse intensities, photon energies, and polarization states in the wide spectral range, as has been recently enabled by the LCLS II variable-gap undulators. Of central importance is the use of time-of-flight to encode electron energy; this relaxes our need to extract photo- and Auger electron spectra from high-resolution imaging sensors. With this technique, we use the representation of energy as a drift time within the $\mu$s window such that the signals are fully compatible with the highest repetition rates of the LCLS II.


\subsection{Diagnostics}

Since our instrument aims at a simultaneous multi-edge characterization of photo- and Auger electron spectra with high energy resolution, it will enable spectral diagnostics in an extended but also redundant, i.e. enhanced collection efficiency scheme. 
This can be particularly powerful in multi-color experiments with large color separations and/or in need of optimal conditions for each of the colors under consideration. 
Not only will it non-invasively complement high-resolution grating-based spectrometer measurements for cumulative characterization of averaged pulses, it will also enable cross referencing of multiple spectrometers with ideally equivalent information, thereby efficiently neglecting systematic effects of trajectory alterations that could compromise the diagnostic validity.
A further advantage is that it can readily be extended to the hard x-ray regime, depending on the photon energies and the targeted absorption edges of available gas targets, enhancing the interdisciplinary use of this instrument within the facility's aims and scientific vision.

The past years have seen the extension of the novel and powerful approach of attosecond angular streaking \cite{Eckle2008} to the time-energy characterization of XFEL pulses \cite{Hartmann2018,MartinRupert}. 
The experimental results provide direct information about the full time–energy distribution of individual stochastic SASE x-ray pulses with attosecond resolution, revealing x-ray pulse and sub-spike duration, intensity substructure, and chirp. 

Angular streaking uses a circularly polarized optical laser to dress target atoms or molecules such that photoionization incurs a rotating energy sweep as a sort of optical carrier clock \cite{Constant1997,Eckle2008}.
Thus, the temporal intensity distribution of the XFEL pulse is mapped to the energy distribution of the streaked photoelectron spectra with the angular distribution providing the necessary temporal characterization of FEL pulses with attosecond resolution \cite{Kienberger2002,Helml2014,Kazansky2016,Siqi2018}. 
In combination with a long wavelength laser field, our instrument will measure directly the x-ray profile \cite{Hartmann2018} via angular streaking, making available X-ray pulse duration measurements where the use of the x-band transverse cavity (XTCAV) is precluded by the high repetition rate of the LCLS-II.
This use applies even at low repetition rate since short x-ray pulses suffer from comparable electron/x-ray slippage such that XTCAV interpretation no longer properly represents bunch dependent lasing.
This need for the angular-streaking alternative of XTCAV will become an increasing concern as the community explores multi-stage amplification schemes \cite{Lutman2016,FreshSlice_Amann,DoubleBunch2017}.

The increasing interest in polarization control motivates the use of MRCOFFEE as a time-resolving x-ray pulse polarimeter.
Given the development of time-dependent polarization states of XFEL pulses as in Ref.~\cite{Sudar2020}, online measurement of the polarization state as in Ref.~\cite{Lutman2016} combined with the attosecond resolution of Refs.~\cite{Nick2018,MartinRupert} portends active use as a spectral and polarization diagnostic during tune up, demonstration, and user use of such novel operation modes. At the LCLS II the possibility of effectively using a reverse taper scheme to deliver circular polarization and of seeding the Delta undulator with the Fresh-slice technique \cite{LutmanFreshSlice2016} will require online measurement of polarization to optimize the cross-polarized undulator beams. Beyond the use for setup of such novel FEL operation modes, this diagnostic capability will allow on-the-fly sorting based on the polarization state of individual shots, bringing also time-dependent polarization into the ``measure-and-sort'' paradigm
On the nanosecond front, our instrument can be used to identify nanosecond separated pulse intensities when running so-called ``multi-bucket'' modes.

Using atomic or molecular photo- and Auger electron spectra with sharp spectral features will allow the identification of time-shifted copies of the spectra, thus revealing the separation and even intensity variation for the multi-bucket pulse train with sub-ns resolution.

\subsection{Science}

Angle resolved photo- and Auger electron spectroscopy (ARPES and ARAES respectively) enable XFELs to tackle a variety of desired scientific objectives across different sample environments and scientific disciplines. 
Building on our history of XFEL based molecular spectroscopy\cite{Cryan10,CryanJPB}, one of the driving goals for this instrument is enabling exquisite soft-x-ray spectroscopic methods.
Targeting simultaneous molecular photoelectron and resonant Auger studies, we have designed for 0.25~eV resolution on both incoming photon energy reconstruction and outgoing electron kinetic energy resolution.
This, the angle resolving spectroscopic capabilities alone of this instrument will serve the broader molecular spectroscopy community \cite{Kiyosi2006,Simon2014,Simon2016} as well as the emerging attosecond resolved studies enabled by angular streaking.

The scope of the associated methodological developments extend well beyond the familiar investigations of the emission characteristics of e.g. crystal lattices, simple atomic or molecular systems, and their orbital structure and dynamics. 
Not only do angle-resolved electron spectra from a dilute gas target non-invasively reveal the state and degree of circular polarization \cite{Lutman2016} and the attosecond time–energy structure of XFEL pulses \cite{Nick2018,Siqi2018,MartinRupert}, but also, it simultaneously grants access to the electron dynamics in small quantum systems.
Depending on the experimental scheme, our instrument can be employed together \textit{in situ} or serially in parallel with experiment specific instruments along the beamline.
The experimental possibilities thus opened up could be applied to the investigation of general light-matter interaction of atomic and molecular targets, ultrafast reaction pathways for photolysis processes in prototypical and eventually complex organic molecules, the study of fundamental charge dynamics e.g. in water to deduce information about DNA damage.
Excitingly, these research opportunities quite generally pave the way to ultrafast x-ray pump-probe experiments at the attosecond frontier \cite{JamesCorrelation,RostPRL2020}, setting the stage for nonlinear attosecond x-ray science that can resolve electron dynamics at their natural time scales \cite{Adrian2021}.

Inspired by the broad applicability of MRCOFFEE to both scientific and diagnostic tasks, we here present a few selected research avenues for fundamental science applications. 
We will specifically spotlight three gas-phase applications that will be enabled, advanced, or leveraged by the XFEL tailored design of the instrument, namely attosecond x-ray physics, polarization control, and molecular-frame spectroscopy.

The novel technique of angular streaking can non-invasively retrieving every shot's time-energy structure and duration with attosecond resolution.
As introduced above, it was demonstrated that even x-ray pump/x-ray probe experiments with sub-fs pulses is feasible by sorting for appropriate delays between SASE spikes. A much more advanced control and delay-resolution has meanwhile been enabled \cite{JamesCorrelation}, bringing time-resolved XFEL x-ray science fully into the attosecond regime \cite{Hartmann2018}.
With the improved design of MRCOFFEE, we will be able to follow up on the recent results of attosecond angular streaking and carry them further. 
In line with the recent demonstration of XLEAP at the LCLS \cite{Duris2019} together with emerging x-ray polarization methods \cite{Sudar2020}, this instrument will fully support the development of new operation modes with high-energy attosecond pulses and innovative measurement schemes. 

A potentially critical concept is that of attoclock ptychography \cite{Feurer2018} in which the application of time-domain ptychography to angular photo-electron streaking is proposed. 
Originally, space-domain ptychography was used to solve the phase problem in crystallography \cite{Hoppe1969} by iteratively reconstructing an image from a series of overlapping far-field diffraction images. 
Attoclock ptychography operates similarly but in the time-frequency domain whereby angular streaking spectra replace the diffraction images of conventional ptychography.
Upon close inspection of Ref.~\cite{Feurer2018}, we find that a compromise in energy resolution of about 0.25\%, e.g. 0.25 eV in 10-75~eV window, achieves an acceptable reconstruction accuracy that only improves with diminishing returns until the spectrometer resolution becomes an order of magnitude better than is achievable for anything other than an angle integrating magnetic bottle spectrometer which itself would preclude angular streaking.
Likewise, increasing the angular sampling above about 6 would also deliver a diminishing return in reconstruction error; the 20 angular samples of the MRCOFFEE instrument are therefore a sufficient compromise for our goal of multiple energy windows.

A similar approach to attoclock ptychography has been dubbed attosecond transient absorption ``spooktroscopy'' for the sake of its relationship to classical ghost-imaging.
Again, drawing inspiration from spatial domain methods for super-resolution, one uses high-resolution ultrafast absorption spectroscopy with broad-bandwidth XFEL pulses \cite{Taran2020,JamesCorrelation} to recover the desired resolution in the time and energy domain. 
Other potential measurements involve the direct time-domain measurement of electronic relaxation dynamics in atoms and molecules as recently demonstrated for the Auger electron decay dynamics in neon atoms \cite{Adrian2021}. 
One of our particular interests lies in temporally resolved measurements of resonant double-core hole processes and decay in both atomic and molecular species with respect to the x-ray pulse intensity substructure at the attosecond scale.
Given the advances in optical pulse shaping for gas-phase molecular alignment \cite{Cryan09} and orientation \cite{VarunMultiKick2014}, we expect broad application of the method of using the rotational coherences induced by optically kicking a molecular rotor for extracting molecular frame emission patterns from lab-frame fluctuations in yield \cite{Hockett2017,Varun2020}.
We therefore expect that the time-energy resolved photo-electron and Auger electron spectroscopy, reconstructed in the molecular frame, as enabled by our instrument will find broad application for the investigation of nonlinear light–matter interactions and coherence effects at x-ray frequencies.\\

The molecular frame angular scattering pattern of ionized electrons, freed from the self same molecule, provides numerous benefits to conventional electron scattering \cite{Rebecca2014,Nakajima2015,Till2020}.
Recent investigations into the attosecond time-resolved photoionization process in molecules has shown that nuclear motion can in fact imprint its effect on the outgoing photo-electron.
Such effects are only visible with sufficient time-energy resolution \cite{Andrei2020}, ideally with reconstruction of the molecular frame emission pattern since the topic of nuclear motion effects on the emission directions are a hot topic related to extent of validity the Franc-Condon principle of molecular transitions \cite{Nandi2020}.
Such recent attosecond photoionization measurements of molecular targets have underlined that molecular ionization is a complex process full of channel-coupling dynamics and rich physics as photo-electrons interact with the molecular orbitals as they rearrange into cationic configurations.
Studying the results of these fundamental scattering processes advances the understanding of electron dynamics in their natural time regime.

Simulations of molecular-frame photoionization show complex angular dependencies that are difficult to measure with conventional detectors due to either a lack of angular resolution or poor energy resolution.
The interpretation of existing measurements \cite{CryanJPB} is also complicated due to partial-wave mixing that results from the stricter symmetry of molecules.
The angular resolution of MRCOFFEE is sufficiently high to measure the lab-frame outgoing partial waves as they fluctuate with the coherent rotational wavepacket.
The molecular frame emission pattern is then reconstructed as per Ref.~\cite{Varun2020}.
Preserving high energy resolution makes this instrument ideal for molecular frame photoionization measurements, enabling partial wave decomposition of photo- and Auger electron wavepackets.
The high energy resolution over a wide energy window for each individual eToF detector is critical to mitigating the the spectral breadth and congestion of the involved molecular cationic states.

Another of the unique opportunities that XFELs offer is the possibility to site specifically address individual constituents of a polyatomic system with ultraintense and ultrashort pulses, thereby enabling an observer-specific vista on nonlinear and ultrafast processes. 
A broad variety of new scientific routes has been initiated, and in the light of rapidly evolving technological and diagnostic advances on both the machine side and the experiment side, rich potential unfolds for the immediate and also longer-term future \cite{Young2018, Ueda2019}. 
A recent technological advance is undulator-based polarization control that enables efficient access to nonlinear and ultrafast dichroic phenomena in the x-ray regime \cite{Allaria2014, Lutman2016}. 
The angular distribution of electrons emitted via circularly polarized light yields a variety of information on the ionizing light as well as access to scientific questions such as dichroic light-matter interaction. 
The circularly polarizing ``Delta'' afterburner undulator was successfully operated at LCLS I and has revealed promising scientific potential for using polarization control for ultrashort XFEL pulses \cite{Durr2016, Hartmann2016Pol}. 
New schemes were recently proposed for enabling coherent x-rays with tunable time-dependent polarization \cite{Sudar2020}. 
Following the success of FERMI in Italy as the first polarization-controlling short-wavelength FEL, LCLS has pioneered polarization control at XFELs and enabled fundamental circular dichroism studies \cite{Hartmann2016Pol}, magnetization studies via x-Ray Magnetic Circular Dichroism (XMCD) \cite{Higley2016}, as well as the first steps into studying dynamics and nonlinear processes in chiral systems with ultrabright x-ray pulses \cite{Schoenlein2015, Ilchen2020}.
Several recent studies from FERMI give an appealing outlook on polarization control as extended to the x-ray regime \cite{Mazza2014, Ilchen2017, Young2018, Ueda2019, DeNinno2020, Carpeggiani2019}. 

The upcoming opportunities include dichroism studies in the nonlinear regime \cite{Ilchen2017, DeNinno2020} for new phenomena of elliptical dichroism \cite{Hofbrucker2018} to non-dipole studies \cite{Ilchen2018, Bachau2019, DeNinno2020}, and the exploration of chirality via photo-electron circular dichroism \cite{Ritchie1976, Wollenhaupt2016, Beaulieu2016, Ilchen2020}.
The unique advantage of the full MRCOFFEE spectrometer array is the potential to retrieve 3D angular-distribution patterns of mixed polarization states while maintaining the high energy resolution afforded by fast anode-mounted MCPs.
The combination of undulator-based polarization control with intense attosecond pulses from XLEAP \cite{Duris2019} can even add valuable perspectives for exploring electron dynamics in chiral systems before and right at the very onset of structural dynamics. 
This novel view on nonlinear and dynamical stereochemistry will strongly require exquisite knowledge of all underlying pulse properties such as the degree of polarization and the single-shot attosecond time-energy structure which underlines the core strengths of the instrument presented here.

\section{Conclusion and Outlook\label{conclusion}}

We have presented the current layout and developments of a new diagnostic and scientific instrument for atomic, molecular and optical physics at the TMO endstation of LCLS-II. 
Some key advances compared to previous instruments targeting angle resolving electron time-of-flight spectroscopy will enable non-invasive and online information about polarization control, spectral composition, and the attosecond time--energy structure of each incoming x-ray pulse. 
One of the core strengths of this novel spectrometer array is an optimized multi-resolution capability that allows for simultaneous investigation, within a single XFEL shot, of multiple different kinetic energy regimes.
This instrument has been also designed for specific compatibility with MHz repetition rates providing the development and demonstration platform for the upcoming technological developments into widely controllable attosecond and novel structured x-ray pulses associated with the active pulse shaping research for the LCLS-II.


\section{Acknowledgements}
This development and research was carried out at the Linac Coherent Light Source (LCLS) at the SLAC National Accelerator Laboratory. 
SLAC National Accelerator Laboratory, is supported by the U.S. Department of Energy, Office of Science, Office of Basic Energy Sciences under Contract No. DE-AC02-76SF00515. 
RNC acknowledges funding through the U.S. Department of Energy, Office of Science, Office of Basic Energy Sciences under Field Work Proposal 100498, ``Enabling long wavelength Streaking for Attosecond X-ray Science'' as well as algorithm development support under Field Work Proposal 100643 ``Actionable Information from Sensor to Data Center.'' 
MI acknowledges funding of the Volkswagen foundation for a Peter-Paul-Ewald Fellowship.
The authors want to thank Timor Osipov, Jeff Aldrich, Justin James, Amore Lopes, Paul Fuoss, Su-Ping Cheng, Alan Conder, James M. Glownia, Micheal Holmes, David Rich, David Fritz, Gregor Hartmann, Markus Braune, Frank Scholz, and Jens Viefhaus, for their constant support and fruitful discussions. 
We thank Richard Davies (MDC Vacuum) for critical input into the design of the Mu-metal chamber. 
We furthermore acknowledge the entire LCLS, LCLS II, L2SI subsystems and SLAC vacuum and machine shops staff for their assistance during the design, construction, and commissioning of the MRCOFFEE instrument. 

\bibliographystyle{unsrt}
\bibliography{bib.bib,Coffee.bib,Wolfi.bib,artof.bib,2dAuger.bib, attosecond.bib}



\end{document}